\newcommand{\mii}{Mg\,\textsc{ii}}
\newcommand{\civ}{C\,\textsc{iv}}

\documentclass{ws-ijmpd}
\usepackage{hyperref}
\usepackage{url}
\usepackage[usenames]{color}
\usepackage[super,compress]{cite}
\begin{document}

\markboth{Park \& Ratra}
{Updated observational constraints on $\phi$CDM dynamical dark energy cosmological models}

%%%%%%%%%%%%%%%%%%%%% Publisher's Area please ignore %%%%%%%%%%%%%%%
%
\catchline{}{}{}{}{}
%
%%%%%%%%%%%%%%%%%%%%%%%%%%%%%%%%%%%%%%%%%%%%%%%%%%%%%%%%%%%%%%%%%%%%

\title{\boldmath Updated observational constraints on $\phi$CDM dynamical dark energy cosmological models}

\author{Chan-Gyung Park}
\address{Division of Science Education and Institute of Fusion Science, Jeonbuk National University,\\
Jeonju 54896, Republic of Korea\\
park.chan.gyung@gmail.com}

\author{Bharat Ratra}
\address{Department of Physics, Kansas State University, 116 Cardwell Hall, Manhattan, KS 66506, USA\\
ratra@ksu.edu}

\maketitle

\begin{history}
\received{Day Month Year}
\revised{Day Month Year}
\end{history}

\begin{abstract}
We present updated observational constraints on the spatially flat $\phi$CDM cosmological model, in which dark energy is described by a minimally coupled scalar field $\phi$ with an inverse power-law potential energy density $V=V_0 \phi^{-\alpha}$. Using a combination of Planck 2018 cosmic microwave background (CMB) temperature, polarization (P18), and lensing power spectra (lensing), along with a comprehensive compilation of non-CMB data including baryon acoustic oscillation, type Ia supernova, Hubble parameter, and growth rate measurements, we analyze parameter constraints of the $\phi$CDM and $\phi$CDM+$A_L$ models where $A_L$ is the CMB lensing consistency parameter. We find that the scalar field parameter $\alpha$, which governs dark energy dynamics, is more tightly constrained by non-CMB data than by CMB data alone. From P18+lensing+non-CMB data, we obtain $\alpha = 0.055 \pm 0.041$ in the $\phi$CDM model and $\alpha = 0.095 \pm 0.056$ in the $\phi$CDM+$A_L$ model, mildly favoring evolving dark energy over a cosmological constant by $1.3\sigma$ and $1.7\sigma$. The estimated Hubble constant is $H_0=67.55_{-0.46}^{+0.53}$ km s$^{-1}$ Mpc$^{-1}$ for P18+lensing+non-CMB data in the $\phi$CDM model, consistent with median statistics and some local determinations, but in tension with other local determinations. The constraints for matter density and clustering amplitude ($\Omega_m = 0.3096 \pm 0.0055$, $\sigma_8 = 0.8013_{-0.0067}^{+0.0077}$) of the flat $\phi$CDM model statistically agree with $\Lambda$CDM model values. Allowing the CMB lensing consistency parameter $A_L$ to vary reduces tensions between CMB and non-CMB data, although we find $A_L = 1.105 \pm 0.037$, $2.8\sigma$ higher than unity, consistent with the excess smoothing seen in Planck data. Model comparison using AIC and DIC shows that the $\phi$CDM model provides a fit comparable to $\Lambda$CDM, with the $\phi$CDM+$A_L$ extension slightly preferred in some cases. Overall, our results indicate that while the $\Lambda$CDM model remains an excellent fit current data leave open the possibility of mildly evolving quintessence-like dynamical dark energy, without a phantom-divide crossing.
\end{abstract}


\ccode{PACS numbers: 98.80.-k}

%\tableofcontents

\section{Introduction}
\label{sec:Introduction} 

The standard spatially flat $\Lambda$CDM cosmological model \cite{Peebles:1984ge}, where dark energy is represented by the cosmological constant $\Lambda$, remains the simplest and most successful framework for describing the large-scale evolution of the universe \cite{Planck:2018nkj, Planck:2018vyg}. This model provides a good fit to a wide range of high- and low-redshift observations, including cosmic microwave background (CMB) anisotropies, baryon acoustic oscillations (BAO), type Ia supernova (SNIa) apparent magnitudes, measurements of the Hubble parameter [$H(z)$], and the growth rate of matter fluctuations ($f\sigma_8$). Despite its empirical success, the $\Lambda$CDM model has unresolved conceptual issues, including the so-called fine-tuning problem associated with the value of the cosmological constant and that it is difficult to accommodate a cosmological constant in the standard model of particle physics \cite{Weinberg:1988cp, Peebles:2002gy}, as well as some potential observational discrepancies \cite{Hu:2023jqc, CosmoVerseNetwork:2025alb}. 

These problems have led to the exploration of models where the dark energy component evolves dynamically. Among these models, widely used parameterizations are those where the dark energy fluid equation of state has a constant value that differs from $w=-1$ (which corresponds to the cosmological constant) or where $w$ varies with redshift $z$ or time. Here $w$ is the ratio of the pressure to the energy density of the dynamical dark energy fluid, and these are known as the X/$w$CDM (with $w$ a constant and $< -1/3$) or $w(z)$CDM parameterizations. It should be noted that the simplest XCDM and $w(z)$CDM parameterizations are physically incomplete.\footnote{The simplest versions of these parameterizations have imaginary speeds of sound, which result in rapidly growing spatial inhomogeneities, and need to be modified to fix this problem.} 

In contrast, dynamical dark energy described by a dynamical scalar field $\phi$ with potential energy density $V(\phi)$ is a physically consistent dynamical dark energy model known as the $\phi$CDM model \cite{Peebles:1987ek, Ratra:1987rm}. With a suitable choice of $V(\phi)$, the energy density of the scalar field, $\rho_\phi$, can be subdominant in the early universe, thus, for example, not affecting standard big bang nucleosynthesis. Near the current epoch, with a suitable choice of $V(\phi)$, $\rho_\phi$ dominates over all other contributions to the cosmic energy budget and drives the observed, late-time, accelerated expansion of the universe. 

One of the simplest forms of the scalar field dynamical dark energy model, with these properties, is described by a minimally coupled scalar field $\phi$ with an inverse power-law potential energy density, $V(\phi)=V_0 \phi^{-\alpha}$ \cite{Peebles:1987ek, Ratra:1987rm}. The parameter $\alpha$ controls the dynamics of the scalar field and of the dark energy density: $\alpha=0$ corresponds to the cosmological constant, while small $\alpha>0$ results in a slowly evolving form of quintessential dark energy. In this model, the scalar field’s energy density evolves along a tracker solution \cite{Peebles:1987ek, Ratra:1987rm}, helping to reduce aspects of the fine-tuning problem and allowing the present accelerated expansion to emerge more naturally from plausible cosmological initial conditions. Since $\phi$CDM predicts a time-varying equation of state, it offers a physically motivated alternative to the purely phenomenological, physically incomplete, XCDM, $w(z)$CDM, or $w_0 w_a$CDM [where $w(z) = w_0 + w_a z/(1+z)$] \cite{Chevallier:2000qy, Linder:2002et} parameterizations.

While there have long been indications that data weakly favor mild dark energy dynamics over a constant cosmological constant, Refs.\ \citen{Sola:2016hnq, Ooba:2017lng, Ooba:2018dzf, Park:2018fxx, SolaPeracaula:2018wwm, Park:2019emi, Khadka:2020vlh, Cao:2020jgu, Khadka:2020hvb, Cao:2021ldv, Cao:2021irf, Dong:2023jtk, VanRaamsdonk:2023ion, VanRaamsdonk:2024sdp, Thompson:2024nxf} and references therein, recent DESI results \cite{DESI:2024mwx, DESI:2025zgx} are more significant, favoring dynamical dark energy over a $\Lambda$ at $\gtrsim2\sigma$ for Ref.\ \citen{DESI:2024mwx} and $2.8\sigma$ for Ref.\ \citen{DESI:2025zgx} from CMB+DESI+SNIa (PantheonPlus) data, and so more interesting, see Refs.\ \citen{RoyChoudhury:2024wri, Huang:2025som, Bansal:2025ipo, Chakraborty:2025syu, Borghetto:2025jrk, Ishiyama:2025bbd, Moghtaderi:2025cns, Urena-Lopez:2025rad, Paliathanasis:2025dcr, Shah:2025ayl, Shlivko:2025fgv, Wolf:2025jed, Mirpoorian:2025rfp, RoyChoudhury:2025dhe, Liu:2025mub, Ye:2025ark, Cheng:2025lod, Mukherjee:2025ytj, vanderWesthuizen:2025iam, Cai:2025mas, Gonzalez-Fuentes:2025lei, Barua:2025ypw, Gialamas:2025pwv, Liu:2025myr, Mishra:2025goj} and references therein. To examine whether dynamical dark energy is favored over a cosmological constant, the DESI analyses \cite{DESI:2024mwx, DESI:2025zgx} used the $w_0w_a$CDM parameterization. Recently, we have used our compilation of CMB and non-CMB (BAO, SNIa, $H(z)$, and $f\sigma_8$) data\cite{deCruzPerez:2024abc} to also constrain the $w_0w_a$CDM parameterization \cite{Park:2024vrw} and found that our data compilation favored dark energy dynamics over a cosmological constant slightly more significantly than did the original DESI analysis \cite{DESI:2024mwx} but less significantly than does the latest DESI analysis.\cite{DESI:2025zgx} Given that the $w_0w_a$CDM parameterization is not physically complete, it is important to use a physically complete model to analyze our (as well as the DESI) data compilation to see whether dynamical dark energy is also indicated in a physically complete model. In this paper we use the $\phi$CDM model in analyses of our data compilation.\footnote{We note here, and discuss in more detail below, that the $w_0w_a$CDM parametrization can accommodate both phantom-like and quintessence-like dark energy dynamics while the simplest $\phi$CDM model we use here can only describe quintessence-like dark energy dynamics.}   

Recent analyses fitting the spatially flat $\phi$CDM model, based on the inverse–power-law $V(\phi)$, to observational data show that allowed dark energy dynamics is at most mild. The best fits favor a small, positive $\alpha$, but remain statistically consistent with $\alpha = 0$ (i.e., $\Lambda$CDM), when using a combination of Planck 2015 CMB data, and BAO, SNe Ia, $H(z)$, and $f\sigma_8$ measurements \cite{Park:2018fxx}.\footnote{Also see Refs.\ \citen{Ooba:2017lng, Ooba:2018dzf} for similar results. For constraints on the $\phi$CDM model from earlier data see Refs.\ \citen{Chae:2004jp, Chen:2004nqb, Samushia:2008fk, Chen:2011ys, Farooq:2012ev, Farooq:2013dra}.}

In our analyses of our data compilation using the $w_0w_a$CDM parameterization, we found that the $\sim2\sigma$ support for dynamical dark energy over a $\Lambda$ did not depend on including DESI data in the analysis (since we find a very similar result using our data compilation which does not contain DESI data) nor did it depend on including Pantheon+ SNIa data \cite{Brout:2022vxf} in our compilation \cite{Park:2024vrw}. However, when we instead allowed the lensing consistency parameter $A_L$\cite{Calabreseetal2008} to vary and also be determined from these data we found the support for dynamical dark energy over a $\Lambda$ decreased to $\sim 1.5\sigma$ with the resulting $A_L$ value being $2.2\sigma$ larger than unity, \cite{Park:2024pew} suggesting that some of the support for dynamical dark energy in the $w_0w_a$CDM parameterization comes from the observed excess smoothing of some of the Planck CMB data multipoles relative to those in the best-fit cosmological model. 

In this paper, we extend previous analyses of dynamical dark energy models by deriving updated parameter constraints on the spatially flat $\phi$CDM model. Our analysis here uses the Planck 2018 CMB temperature, polarization, and lensing measurements \cite{Planck:2018nkj,Planck:2018vyg} in combination with a large, mutually-consistent non-CMB dataset, which includes BAO, SNIa, $H(z)$, and $f\sigma_8$ observations \cite{deCruzPerez:2024abc}. We also examine the extended $\phi$CDM+$A_L$ model, allowing the CMB lensing amplitude parameter $A_L$ to vary, to determine whether we find the same effect we saw in the XCDM, $w_0w_a$CDM, and $w(z)$CDM parameterizations \cite{deCruzPerez:2024abc, Park:2024pew, Park:2025azv, deCruzPerez:2026mkg}.

For the largest data set we use here (P18+lensing+non-CMB) we find $\alpha = 0.055 \pm 0.041$ ($\alpha <0.133$, 95\% upper limit) in the $\phi$CDM model and $\alpha = 0.095 \pm 0.056$ ($\alpha < 0.196$, 95\% upper limit) in the $\phi$CDM+$A_L$ model, both of which are consistent with a $\Lambda$ ($\alpha=0$), but both of which allow mild quintessence-like dark energy dynamics, at $1.3\sigma$ and $1.7\sigma$ significance respectively. Allowing the CMB lensing amplitude consistency parameter $A_L$ to vary reduces tensions between CMB data and non-CMB data constraints, although we find $A_L = 1.105 \pm 0.037$, $2.8\sigma$ higher than unity, consistent with the excess smoothing seen in Planck data. Goodness-of-fit model comparisons show that the $\phi$CDM model provides a fit comparable to the $\Lambda$CDM model, with the $\phi$CDM+$A_L$ model extension slightly preferred in some cases. 

Interestingly, unlike what we find in the XCDM, $w_0w_a$CDM, and $w(z)$CDM parameterizations for the same data compilation we use here, \cite{deCruzPerez:2024abc, Park:2024pew, Park:2025azv} in the $\phi$CDM model when $A_L$ is allowed to vary and be measured from these data the support for dark energy dynamics increases relative to the $A_L = 1$ case. Overall, our results indicate that the $\Lambda$CDM model remains an excellent fit but leave open the possibility of mildly evolving quintessence-like dynamical dark energy and show that these data are reasonably consistent with dark energy dynamics without a phantom-divide crossing.

The structure of our paper is as follows. In Sec.~\ref{sec:Data} we describe the datasets used. In Sec.~\ref{sec:Methods} we outline the $\phi$CDM model and our analysis methodology. In Sec.~\ref{sec:ResultsandDiscussion} we present the parameter constraints and model comparisons. Finally, in Sec.~\ref{sec:Conclusion} we summarize our conclusions and the implications for dark energy dynamics.

\section{Data}
\label{sec:Data}

We use CMB and non-CMB measurements to constrain $\phi$CDM model cosmological parameters. The data we use here are described in detail in Sec.\ II of Ref.\ \citen{deCruzPerez:2024abc} and summarized below. In our analyses we account for all known data covariances. 

The CMB data we use are the Planck 2018 TT,TE,EE+lowE (P18) CMB temperature and polarization power spectra alone as well as jointly with the Planck lensing potential (lensing) power spectrum \cite{Planck:2018nkj,Planck:2018vyg}. 

The non-CMB data we use is the non-CMB (new) data compilation of Ref.\ \citen{deCruzPerez:2024abc} comprised of 

\begin{itemize}

\item 16 BAO measurements that span $0.122 \le z \le 2.334$ and are listed in Table I of Ref.\ \citen{deCruzPerez:2024abc}. These include low-redshift data from the 6dFGS and SDSS MGS surveys, intermediate-redshift data from BOSS galaxies ($z = 0.38$ and $0.51$), eBOSS LRG ($z = 0.698$) and DES year 3 ($z=0.835$) and high-redshift data from eBOSS quasars ($z=1.38$) and the Ly$\alpha$ forest ($z = 2.334$). In addition to several distances ($D_V$, $D_M$, and $D_H$) these also include redshift-space distortion (RSD)--derived growth rate measurements $f\sigma_{8}$. Full covariance matrices are used for correlated BOSS, eBOSS LRG and quasars, and Ly$\alpha$ data. In this work we do not use DESI BAO data \cite{DESI:2025zgx} to remain independent of DESI and consistent with our earlier analyses.

\item The 1590 SNIa apparent magnitude vs.\ redshift measurements subset of the Pantheon+ compilation \cite{Brout:2022vxf} that includes only SNIa with $z > 0.01$ to minimize contamination from local peculiar velocities. This dataset covers a wide redshift interval, $0.01016 \le z \le 2.26137$, and includes both statistical and systematic uncertainties. The absolute magnitude of SNIa is treated as a nuisance parameter and marginalized over.

\item 32 Hubble parameter [$H(z)$] data points that span $0.070 \le z \le 1.965$,  primarily derived from cosmic chronometers, and are listed in Table 1 of Ref.\ \citen{Cao:2023eja} and in Table II of Ref.\ \citen{deCruzPerez:2024abc}.   

\item 9 additional (non-BAO) growth rate ($f\sigma_8$) measurements that
span $0.013 \le z \le 1.36$, listed in Table III of Ref.\ \citen{deCruzPerez:2024abc}.

\end{itemize}

In total we utilize five individual and combined sets of data sets to constrain the flat $\phi$CDM model: P18, P18+lensing, non-CMB, P18+non-CMB, and P18+lensing+non-CMB data.

\section{Methods}
\label{sec:Methods}

In this work we consider the flat $\phi$CDM model with a minimally coupled dynamical dark energy scalar field $\phi$ with an inverse power-law potential energy density, \cite{Peebles:1987ek,Ratra:1987rm}
\begin{equation}
    V(\phi)=\frac{V_0}{\phi^\alpha},
\end{equation}
where $\alpha$ is a non-negative constant and $\alpha=0$ corresponds to the cosmological constant dark energy. 

We evolve the $\phi$CDM model universe by accounting for radiation, baryonic and cold dark matter, neutrinos, and the scalar field dark energy component, and compare $\phi$CDM model predictions to observations to constrain $\phi$CDM model parameter values. We assume that the scalar field is directly coupled only to the gravitational field. We evolve the scalar field by considering the evolution of both a spatially homogeneous background component and a spatially inhomogeneous linear perturbation variable; see Refs.\ \citen{Hwang:2001uaa,Hwang:2001qk} for the evolution equations for the linear perturbations in the presence of the scalar field. When evolving the homogeneous background scalar field, we use the initial conditions of Ref.\ \citen{Peebles:1987ek} at a scale factor $a_i = 10^{-10}$, which places the homogeneous background scalar field on the attractor/tracker solution.\cite{Peebles:1987ek,Ratra:1987rm,Pavlov:2013nra} When evolving the spatially inhomogeneous scalar field perturbations, we choose the initial values of the scalar field perturbation $\delta\phi$ and its time derivative to vanish ($\delta\phi=0=\delta\phi^{\prime}$) in the CDM-comoving gauge, which is equivalent to the synchronous gauge with the residual gauge modes appropriately removed \cite{Hwang:2001qk}. The evolution of perturbations of the scalar field have a tracking behavior and generally does not depend on the choice of initial conditions \cite{Ratra:1987rm, Brax:2000yb}.

For the inverse power-law scalar field potential energy density, the background evolution of the scalar field is obtained by numerically solving the equation of motion of the scalar field,
\begin{equation}
    \phi^{\prime\prime}+ \left(1+\frac{\dot H}{H^2} \right) \phi^\prime - \hat{V}_0 \alpha \phi^{-\alpha-1} \left(\frac{H_0}{H}\right)^2=0,
\end{equation}
where $\phi^\prime \equiv d\phi /d\ln a$, $H={\dot a}/a$, $\hat{V}_0 \equiv V_0 /H_0^2$, the time derive $d/dt$ is denoted by an overdot, and $H_0$ is the Hubble constant. The Hubble parameter $H(a)$ can be written as 
\begin{equation}
\begin{split}
    & \left( \frac{H}{H_0} \right)^2  =  \\ 
    & \frac{6}{6-(\phi^\prime)^2} 
    \bigg[\Omega_\gamma a^{-4} + (\Omega_b + \Omega_c) a^{-3} + 
        \Omega_\nu (a) + \frac{1}{3}\hat{V}_0 \phi^{-\alpha}\bigg],
\end{split}
\end{equation}
where $a$ is the cosmic scale factor normalized to unity at present, $\Omega_\gamma$, $\Omega_b$, and $\Omega_c$ are the CMB photon, baryon, and CDM density parameter at the present epoch, respectively. $\Omega_\nu (a)$ denotes the contribution from massless and massive neutrinos. $\Omega_\gamma$ and $\Omega_\nu (a)$ are determined from the present CMB temperature $T_0 = 2.7255$ K, the effective number of neutrino species $N_\textrm{eff} = 3.046$, with a single massive neutrino species of mass 0.06 eV. Here we have chosen units such that $8\pi G \equiv 1$.

The analysis methods we use are described in Sec.\ III of Refs.\ \citen{deCruzPerez:2024abc}. A brief summary follows.

We use the \texttt{CAMB}/\texttt{COSMOMC} program (October 2018 version) \cite{Challinor:1998xk,Lewis:1999bs,Lewis:2002ah} to determine observational constraints on $\phi$CDM cosmological model parameters, and for model comparison. \texttt{CAMB} is used to compute the evolution of $\phi$CDM model spatial inhomogeneities and to determine $\phi$CDM model theoretical predictions which depend on cosmological parameters. \texttt{COSMOMC} uses the Markov chain Monte Carlo (MCMC) method to compare these $\phi$CDM model predictions to observational data and determine cosmological parameter likelihoods. The MCMC chains are assumed to have converged when the Gelman and Rubin $R$ statistic satisfies $R-1 < 0.01$ (but see below for two exceptions). We use \texttt{GetDist} code \cite{Lewis:2019xzd} to assess whether the MCMC chains have converged and to compute the average values, confidence intervals, and likelihood distributions of model parameters from the converged MCMC chains. 

In the standard flat $\Lambda$CDM model it is conventional to chose the six primary cosmological parameters to be the current value of the physical baryonic matter density parameter $\Omega_b h^2$, the current value of the physical CDM density parameter $\Omega_c h^2$, the sound horizon angular size at recombination $100\theta_{\text{MC}}$, the reionization optical depth $\tau$, the primordial scalar-type perturbation power spectral index $n_s$, and the power spectrum amplitude $\ln(10^{10}A_s)$, where $h$ is $H_0$ in units of 100 km s$^{-1}$ Mpc$^{-1}$. In the flat $\phi$CDM model we follow Ref.\ \citen{Park:2018fxx} and choose $H_0$ as a primary cosmological parameter instead of $100\theta_{\text{MC}}$. In the $\phi$CDM model considered here, $\alpha$, characterizing the dynamics of dark energy, is adopted as the seventh primary cosmological parameter. We also consider the flat $\phi$CDM+$A_L$ model where the lensing consistency parameter $A_L$ \cite{Calabreseetal2008} is the eighth primary cosmological parameter allowed to vary and be determined from observational data.

We assume flat priors for the primary cosmological parameters, non-zero over: $0.005 \le \Omega_b h^2 \le 0.1$, $0.001 \le \Omega_c h^2 \le 0.99$, $0.5 \le 100\theta_\textrm{MC} \le 10$ (only in the $\Lambda$CDM model), $0.01 \le \tau \le 0.8$, $0.8 \le n_s \le 1.2$, $1.61 \le \ln(10^{10} A_s) \le 3.91$, $0.2 \le h \le 1$ (only in the $\phi$CDM(+$A_L$) models), and $0 \le A_L \le 10$ (only in the $\phi$CDM+$A_L$ models). In the $\phi$CDM model, for the dynamical dark energy parameter we assume a flat prior non-zero over $0 \le \alpha \le 10$. In the $\phi$CDM+$A_L$ model, where the $A_L$ parameter is freely varying, for P18 as well as P18+lensing data, such a wide prior on $\alpha$ leads to a second observationally favored region with $\alpha$ greater than 5, $H_0$ less than 60 km s$^{-1}$ Mpc$^{-1}$, and $\Omega_m$ greater than 0.5, in addition to the more conventional favored region close to the standard $\Lambda$CDM model. Because of the two favored regions for P18 and P18+lensing data, convergence of the MCMC chains was much slower than for the other data sets. To address these issues, we first ran additional analyses with a restricted flat $\alpha$ prior non-zero over $0 \le \alpha \le 5$. In this case convergence was also, but not as, slow, but we halted the runs at $R-1 < 0.0235$ (P18) and at $R-1 < 0.0296$ (P18+lensing) before moving on to a more restricted flat $\alpha$ prior non-zero only over $0 \le \alpha \le 2$. With this narrower prior, convergence improved, reaching $R-1 < 0.01$ for all data sets. In the following we present $\phi$CDM+$A_L$ model results for both restricted $\alpha$ priors, but place our main focus on the case $0 \le \alpha \le 2$.

When we estimate parameters using non-CMB data, we fix the values of $\tau$ and $n_s$ to those obtained from P18 data (since these parameters cannot be determined solely from non-CMB data) and constrain the other cosmological parameters. Additionally, in the $\phi$CDM(+$A_L$) models we also present constraints on three derived parameters: $100\theta_{\text{MC}}$, the current value of the non-relativistic matter density parameter $\Omega_m$, and the amplitude of matter fluctuations $\sigma_8$.

For the spatially-flat tilted $\phi$CDM(+$A_L$) models the primordial scalar-type energy density perturbation power spectrum we use is, \cite{Lucchin:1984yf, Ratra:1989uv, Ratra:1989uz},
\begin{equation}
    P_\delta (k) = A_s \left( \frac{k}{k_0} \right)^{n_s}.
\label{eq:powden-flat}
\end{equation}

To quantify how relatively well the $\phi$CDM(+$A_L$) models fit the different data sets under study, we use differences in the Akaike information criterion ($\Delta$AIC) and the deviance information criterion ($\Delta$DIC) between the information criterion (IC) values for the flat dynamical dark energy $\phi$CDM(+$A_L$) models and the flat $\Lambda$CDM model. See Sec.\ III of Ref.\ \citen{deCruzPerez:2024abc}, and references therein, for a more detailed description of these criteria. According to the conventional Jeffreys' scale, when $-2 \leq \Delta\textrm{IC}<0$ there is weak evidence in favor of the model under study, when $-6 \leq \Delta\textrm{IC} < -2$ there is positive evidence, when $-10\leq\Delta\textrm{IC} < -6$ there is strong evidence, and when $\Delta\textrm{IC} < -10$ there is very strong evidence in favor of the model under study relative to the standard tilted flat $\Lambda$CDM model. If the $\Delta\textrm{IC}$ values are positive the $\Lambda$CDM model is favored over the model under study. 

Prior to jointly analyzing two data sets in a given model we need to determine how consistent the cosmological parameter constraints from the individual data sets are in that model. To determine (in)consistency we consider two different statistical estimators. The first one, $\log_{10}\mathcal{I}$, makes use of DIC values, see Ref.\ \citen{Joudaki:2016mvz} and Sec.\ III of Ref.\ \citen{deCruzPerez:2024abc}. Positive values, $\log_{10}\mathcal{I}>0$, indicate consistency, while negative values, $\log_{10}\mathcal{I}<0$, mean that the two data sets are inconsistent. According to the conventional Jeffreys' scale, the degree of consistency or inconsistency between the two data sets is said to be substantial when $\lvert \log_{10}\mathcal{I} \rvert >0.5$, strong when $\lvert \log_{10}\mathcal{I} \rvert >1$, and decisive when $\lvert \log_{10}\mathcal{I} \rvert >2$ \cite{Joudaki:2016mvz}. The second estimator we use is the tension probability $p$ and corresponding Gaussian approximation "sigma value" $\sigma$, see Refs.\ \citen{Handley:2019pqx, Handley:2019wlz, Handley:2019tkm} and Sec.\ III of Ref.\ \citen{deCruzPerez:2024abc}. In the Gaussian approximation, $p=0.05$ approximately corresponds to a $2\sigma$ Gaussian deviation, while $p=0.003$ corresponds to a 3$\sigma$ Gaussian deviation.

%%%%%%%%%%%%%  TILTED FLAT phiCDM %%%%%%%%%%%%%% combined table

\begin{table}[htbp]
\tbl{Mean and 68\% (or 95\%) confidence limits of flat $\phi$CDM model parameters
        from non-CMB, P18, P18+lensing, P18+non-CMB, and P18+lensing+non-CMB data. $H_0$ has units of km s$^{-1}$ Mpc$^{-1}$. }
{\begin{tabular}{@{}cccccc@{}} \toprule
  Parameter      &  Non-CMB      & P18        &  P18+lensing     &  P18+non-CMB         & P18+lensing+non-CMB     \\
\colrule
  $\Omega_b h^2$                & $0.0319^{+0.0039}_{-0.0046}$ & $0.02234 \pm 0.00015$       & $0.02235 \pm 0.00015$       &  $0.02253 \pm 0.00014$       &  $0.02252 \pm 0.00013$  \\
  $\Omega_c h^2$                & $0.0976^{+0.0062}_{-0.0096}$ & $0.1203 \pm 0.0014$         & $0.1203 \pm 0.0012$         &  $0.11781 \pm 0.00096$       &  $0.11808 \pm 0.00089$  \\
  $H_0$                         & $69.7 \pm 2.5$               & $64.2^{+3.1}_{-1.3}$        & $64.7^{+2.6}_{-1.1}$        &  $67.57^{+0.56}_{-0.48}$     &  $67.55^{+0.53}_{-0.46}$       \\
  $\tau$                        & $0.0546$                     & $0.0546 \pm 0.0078$         & $0.0551 \pm 0.0074$         &  $0.0564^{+0.0072}_{-0.0081}$&  $0.0588^{+0.0066}_{-0.0077}$    \\
  $n_s$                         & $0.9645$                     & $0.9645 \pm 0.0044$         & $0.9644 \pm 0.0041$         &  $0.9703 \pm 0.0038$         &  $0.9695 \pm 0.0037$    \\
  $\ln(10^{10} A_s)$            & $3.63 \pm 0.19$              & $3.046 \pm 0.016$           & $3.047 \pm 0.014$           &  $3.043 \pm 0.016$           &  $3.050^{+0.013}_{-0.015}$      \\
  $\alpha$                      & $0.52^{+0.17}_{-0.15}$       & $0.31 \pm 0.30$ ($<0.925$)  & $0.25 \pm 0.23$ ($<0.717$)  &  $0.063 \pm 0.044$ ($<0.146$)&  $0.055 \pm 0.041$ ($<0.133$)    \\
 \colrule
  $100\theta_\textrm{MC}$       & $1.0190^{+0.0081}_{-0.011}$  & $1.04071 \pm 0.00031$       & $1.04071 \pm 0.00031$       &  $1.04100 \pm 0.00029$       &  $1.04096 \pm 0.00029$  \\
  $\Omega_m$                    & $0.2676^{+0.0085}_{-0.013}$   & $0.349^{+0.013}_{-0.035}$   & $0.343^{+0.012}_{-0.030}$   &  $0.3089 \pm 0.0058$         &  $0.3096 \pm 0.0055$    \\
  $\sigma_8$                    & $0.826 \pm 0.025$            & $0.783^{+0.030}_{-0.013}$   & $0.788^{+0.024}_{-0.010}$   &  $0.7971^{+0.0093}_{-0.0083}$&  $0.8013^{+0.0077}_{-0.0067}$    \\
 \colrule
  $\chi_{\textrm{min}}^2$       & $1458.38$                    & $2765.79$                   & $2774.83$                   & $4240.02$                    &  $4249.27$              \\
  $\Delta\chi_{\textrm{min}}^2$ & $-11.55$                     & $-0.01$                     & $+0.12$                     & $-0.22 $                     &  $+0.01$              \\
  $\textrm{DIC}$                & $1467.96$                    & $2821.36$                   & $2829.61$                   & $4293.90$                    &  $4302.89$              \\
  $\Delta\textrm{DIC}$          & $-10.15$                     & $+3.43$                     & $+3.16$                     & $+1.57$                      &  $+1.69$                \\
  $\textrm{AIC}$                & $1468.38$                    & $2821.79$                   & $2830.83 $                  & $4296.02$                    &  $4305.27$              \\
  $\Delta\textrm{AIC}$          & $-9.55$                      & $+1.99$                     & $+2.12$                     & $+1.78$                      &  $+2.01$                \\
\botrule
\end{tabular}\label{tab:results_flat_phiCDM}}
\end{table}

%%%%%%%%%%%%%  TILTED FLAT phiCDM (alpha < 5) %%%%%%%%%%%%%% combined table

\begin{table}[htbp]
\tbl{Mean and 68\% (or 95\%) confidence limits of flat $\phi$CDM$+A_L$ model parameters
        from non-CMB, P18, P18+lensing, P18+non-CMB, and P18+lensing+non-CMB data. $H_0$ has units of km s$^{-1}$ Mpc$^{-1}$. For the P18 and P18+lensing cases the prior $\alpha \le 5$ was applied.}
{\begin{tabular}{@{}cccccc@{}} \toprule
  Parameter       &  Non-CMB      & P18 ($\alpha < 5$)    & P18+lensing ($\alpha <5$)   & P18+non-CMB     & P18+lensing+non-CMB     \\
\colrule
  $\Omega_b h^2$                & $0.0319^{+0.0039}_{-0.0046}$ & $0.02262 \pm 0.00018$       & $0.02253 \pm 0.00017$       & $0.02272 \pm 0.00015$       & $0.02264 \pm 0.00014$  \\
  $\Omega_c h^2$                & $0.0976^{+0.0062}_{-0.0096}$ & $0.1176 \pm 0.0016$         & $0.1180 \pm 0.0015$         & $0.1165 \pm 0.0010$         & $0.1166 \pm 0.0010$  \\
  $H_0$                         & $69.7 \pm 2.5$               & $57.6 \pm 7.0$              & $59.6 \pm 6.0$              & $67.77 \pm 0.58$            & $67.72^{+0.61}_{-0.54}$       \\
  $\tau$                        & $0.0546$                     & $0.0480 \pm 0.0089$         & $0.0480 \pm 0.0084$         & $0.0501^{+0.0085}_{-0.0074}$& $0.0500^{+0.0085}_{-0.0076}$    \\
  $n_s$                         & $0.9645$                     & $0.9726 \pm 0.0050$         & $0.9705 \pm 0.0048$         & $0.9748 \pm 0.0040$         & $0.9737 \pm 0.0040$    \\
  $\ln(10^{10} A_s)$            & $3.63 \pm 0.19$              & $3.026^{+0.018}_{-0.016}$   & $3.025^{+0.018}_{-0.016}$   & $3.027^{+0.018}_{-0.016}$   & $3.027^{+0.018}_{-0.016}$      \\
  $A_L$                         & $1$                          & $1.33^{+0.12}_{-0.14}$      & $1.160^{+0.060}_{-0.11}$    & $1.224 \pm 0.064$           & $1.105 \pm 0.037$   \\
  $\alpha$                      & $0.52^{+0.17}_{-0.15}$       & $2.1 \pm 1.6$ ($<4.46$)     & $1.4 \pm 1.3$ ($<3.95$)     & $0.099 \pm 0.056$ ($<0.200$)& $0.095 \pm 0.056$ ($<0.196$)    \\
\colrule
  $100\theta_\textrm{MC}$       & $1.0190^{+0.0081}_{-0.011}$  & $1.04102 \pm 0.00033$       & $1.04094 \pm 0.00032$       & $1.04113 \pm 0.00030$       & $1.04111 \pm 0.00030$  \\
  $\Omega_m$                    & $0.2676^{+0.0085}_{-0.013}$  & $0.443 \pm 0.098$           & $0.410 \pm 0.084$           & $0.3047 \pm 0.0059$         & $0.3052 \pm 0.0059$    \\
  $\sigma_8$                    & $0.826 \pm 0.025$            & $0.676 \pm 0.079$           & $0.705 \pm 0.069$           & $0.783^{+0.011}_{-0.0098}$  & $0.783^{+0.011}_{-0.0097}$    \\
 \colrule
  $\chi_{\textrm{min}}^2$       & $1458.38$                    & $2761.41$                   & $2773.15$                   & $4225.26$                   & $4240.90$              \\
  $\Delta\chi_{\textrm{min}}^2$ & $-11.55$                     & $-4.39$                     & $-1.56$                     & $-14.98$                    & $-8.36$              \\
  $\textrm{DIC}$                & $1467.96$                    & $2810.72$                   & $2827.26$                   & $4283.39$                   & $4297.30$              \\
  $\Delta\textrm{DIC}$          & $-10.15$                     & $-7.21$                     & $+0.81$                     & $-8.94$                     & $-3.90$                \\
  $\textrm{AIC}$                & $1468.38$                    & $2819.41$                   & $2831.15 $                  & $4283.26$                   & $4298.90$              \\
  $\Delta\textrm{AIC}$          & $-9.55$                      & $-0.39$                     & $+2.44$                     & $-10.98$                    & $-4.36$                \\
\botrule
\end{tabular}\label{tab:results_flat_phiCDM_a5}}
\end{table}

%%%%%%%%%%%%%  TILTED FLAT phiCDM (alpha < 2) %%%%%%%%%%%%%% combined table

\begin{table}
\tbl{Mean and 68\% (or 95\%) confidence limits of flat $\phi$CDM$+A_L$ model parameters
        from non-CMB, P18, P18+lensing, P18+non-CMB, and P18+lensing+non-CMB data. $H_0$ has units of km s$^{-1}$ Mpc$^{-1}$. For the P18 and P18+lensing cases the prior $\alpha \le 2$ was applied.}
{\begin{tabular}{@{}cccccc@{}} \toprule
  Parameter                     &  Non-CMB                     & P18 ($\alpha < 2$)          & P18+lensing ($\alpha <2$)   & P18+non-CMB                 & P18+lensing+non-CMB     \\
 \colrule
  $\Omega_b h^2$                & $0.0319^{+0.0039}_{-0.0046}$ & $0.02260 \pm 0.00017$       & $0.02253 \pm 0.00017$       & $0.02272 \pm 0.00015$       & $0.02264 \pm 0.00014$  \\
  $\Omega_c h^2$                & $0.0976^{+0.0062}_{-0.0096}$ & $0.1180 \pm 0.0016$         & $0.1182 \pm 0.0015$         & $0.1165 \pm 0.0010$         & $0.1166 \pm 0.0010$  \\
  $H_0$                         & $69.7 \pm 2.5$               & $61.5^{+3.1}_{-5.3}$        & $62.3^{+4.8}_{-3.1}$        & $67.77 \pm 0.58$            & $67.72^{+0.61}_{-0.54}$       \\
  $\tau$                        & $0.0546$                     & $0.0489^{+0.0082}_{-0.0073}$& $0.0485^{+0.0085}_{-0.0075}$& $0.0501^{+0.0085}_{-0.0074}$& $0.0500^{+0.0085}_{-0.0076}$    \\
  $n_s$                         & $0.9645$                     & $0.9714 \pm 0.0049$         & $0.9700 \pm 0.0048$         & $0.9748 \pm 0.0040$         & $0.9737 \pm 0.0040$    \\
  $\ln(10^{10} A_s)$            & $3.63 \pm 0.19$              & $3.028^{+0.018}_{-0.015}$   & $3.027^{+0.018}_{-0.016}$   & $3.027^{+0.018}_{-0.016}$   & $3.027^{+0.018}_{-0.016}$      \\
  $A_L$                         & $1$                          & $1.237^{+0.072}_{-0.083}$   & $1.113^{+0.047}_{-0.059}$   & $1.224 \pm 0.064$           & $1.105 \pm 0.037$   \\
  $\alpha$                      & $0.52^{+0.17}_{-0.15}$       & $0.83 \pm 0.57$ ($<1.83$)   & $0.69 \pm 0.53$ ($<1.72$)   & $0.099 \pm 0.056$ ($<0.200$)& $0.095 \pm 0.056$ ($<0.196$)    \\ 
 \colrule
  $100\theta_\textrm{MC}$       & $1.0190^{+0.0081}_{-0.011}$  & $1.04097 \pm 0.00033$       & $1.04093 \pm 0.00032$       & $1.04113 \pm 0.00030$       & $1.04111 \pm 0.00030$  \\
  $\Omega_m$                    & $0.2676^{+0.0085}_{-0.013}$  & $0.378^{+0.042}_{-0.061}$   & $0.368^{+0.032}_{-0.059}$   & $0.3047 \pm 0.0059$         & $0.3052 \pm 0.0059$    \\
  $\sigma_8$                    & $0.826 \pm 0.025$            & $0.730 \pm 0.041$           & $0.739^{+0.053}_{-0.032}$   & $0.783^{+0.011}_{-0.0098}$  & $0.783^{+0.011}_{-0.0097}$    \\
\colrule
  $\chi_{\textrm{min}}^2$       & $1458.38$                    & $2756.65$                   & $2772.16$                   & $4225.26$                   & $4240.90$              \\
  $\Delta\chi_{\textrm{min}}^2$ & $-11.55$                     & $-9.15$                     & $-2.55$                     & $-14.98$                    & $-8.36$              \\
  $\textrm{DIC}$                & $1467.96$                    & $2812.61$                   & $2826.47$                   & $4283.39$                   & $4297.30$              \\
  $\Delta\textrm{DIC}$          & $-10.15$                     & $-5.32$                     & $+0.02$                     & $-8.94$                     & $-3.90$                \\
  $\textrm{AIC}$                & $1468.38$                    & $2812.65$                   & $2828.16 $                  & $4283.26$                   & $4298.90$              \\
  $\Delta\textrm{AIC}$          & $-9.55$                      & $-7.15$                     & $-0.55$                     & $-10.98$                    & $-4.36$                \\
\botrule
\end{tabular}\label{tab:results_flat_phiCDM_a2}}
\end{table}

%%%%%%%%%%%%%%%%%%%%%%%%%%%%%%%%%%%%%%%%%%%%%%%%%%%%%%%%%%%%%%%%%%%%%%%%%%%%%%%%%%%%%%%%%%%%%%%%
\begin{figure*}[htbp]
\centering
\mbox{\includegraphics[width=127mm]{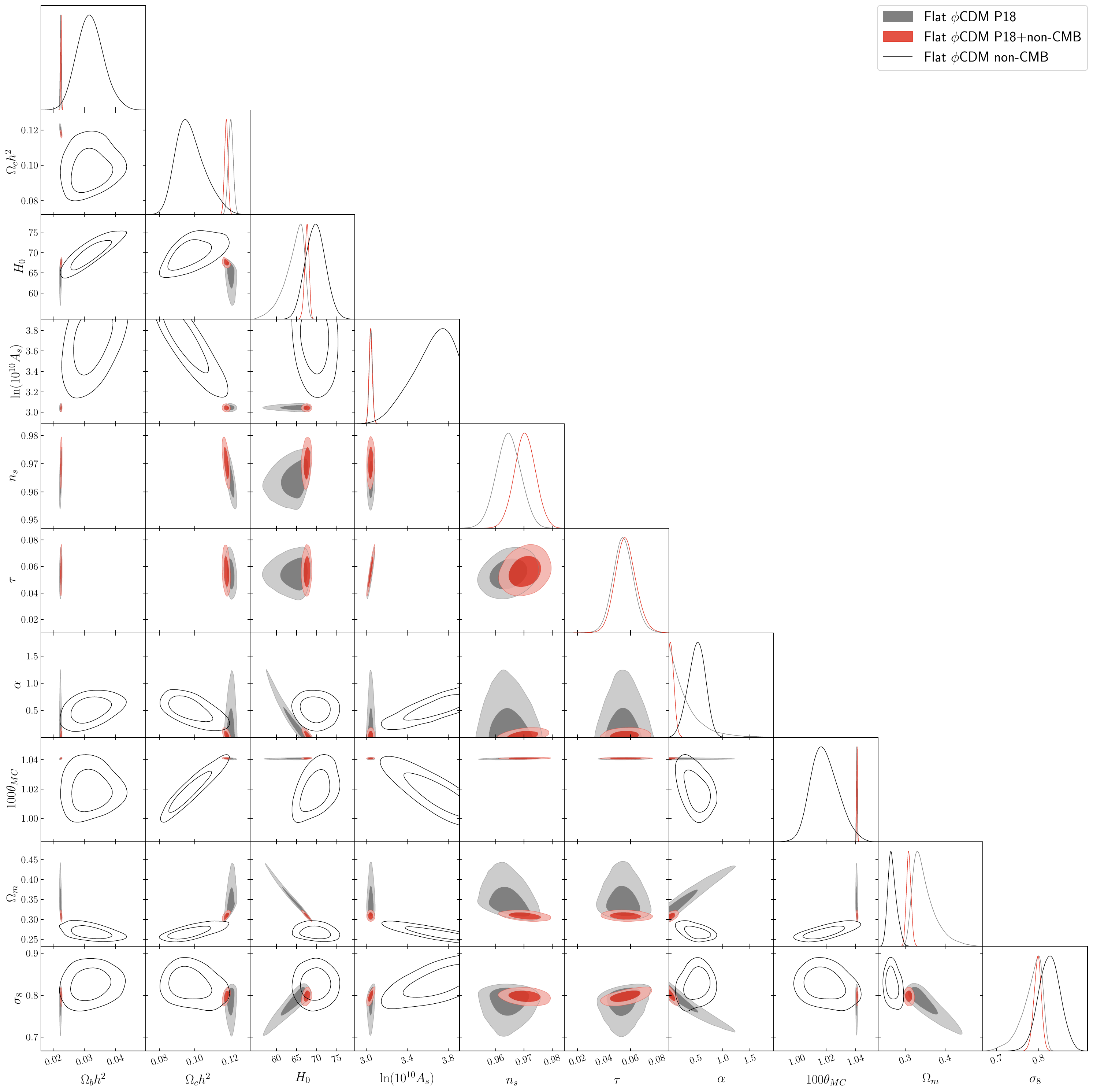}}
        \caption{One-dimensional likelihoods and 1$\sigma$ and $2\sigma$ likelihood confidence contours of flat $\phi$CDM model parameters favored by non-CMB (solid curves), P18 (grey), and P18+non-CMB data sets (red contours). For P18 and P18+non-CMB data cases, we include $\tau$ and $n_s$, which are fixed in the non-CMB data analysis. $H_0$ has units of km s$^{-1}$ Mpc$^{-1}$.      
}
\label{fig:flat_phiCDM_P18_nonCMB23v2_nstau}
\end{figure*}

%%%%%%%%%%%%%%%%%%%%%%%%%%%%%%%%%%%%%%%%%%%%%%%%%%%%%%%%%%%%%%%%%%%%%%%%%%%%%%%%%%%%%%%%%%%%%%%%
\begin{figure*}[htbp]
\centering
\mbox{\includegraphics[width=127mm]{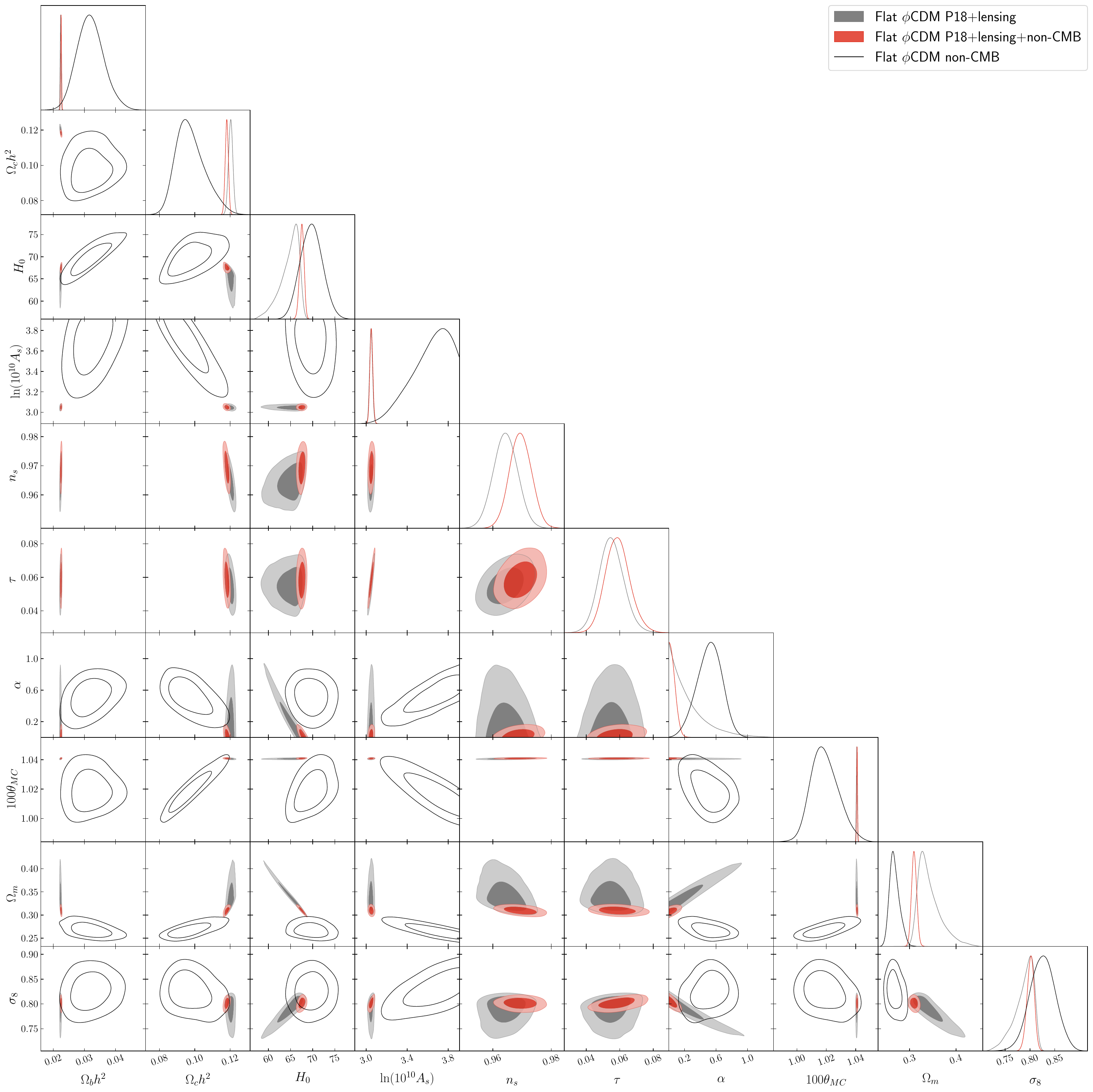}}
        \caption{One-dimensional likelihoods and 1$\sigma$ and $2\sigma$ likelihood confidence contours of flat $\phi$CDM model parameters favored by non-CMB (solid curves), P18+lensing (grey), P18+lensing+non-CMB data sets (red contours). For P18 and P18+lensing+non-CMB cases, we include $\tau$ and $n_s$, which are fixed in the non-CMB data analysis. $H_0$ has units of km s$^{-1}$ Mpc$^{-1}$.
}
\label{fig:flat_phiCDM_P18_lensing_nonCMB23v2_nstau}
\end{figure*}

%%%%%%%%%%%%%%%%%%%%%%%%%%%%%%%%%%%%%%%%%%%%%%%%%%%%%%%%%%%%%%%%%%%%%%%%%%%%%%%%%%%%%%%%%%%%%%%%
\begin{figure*}[htbp]
\centering
\mbox{\includegraphics[width=127mm]{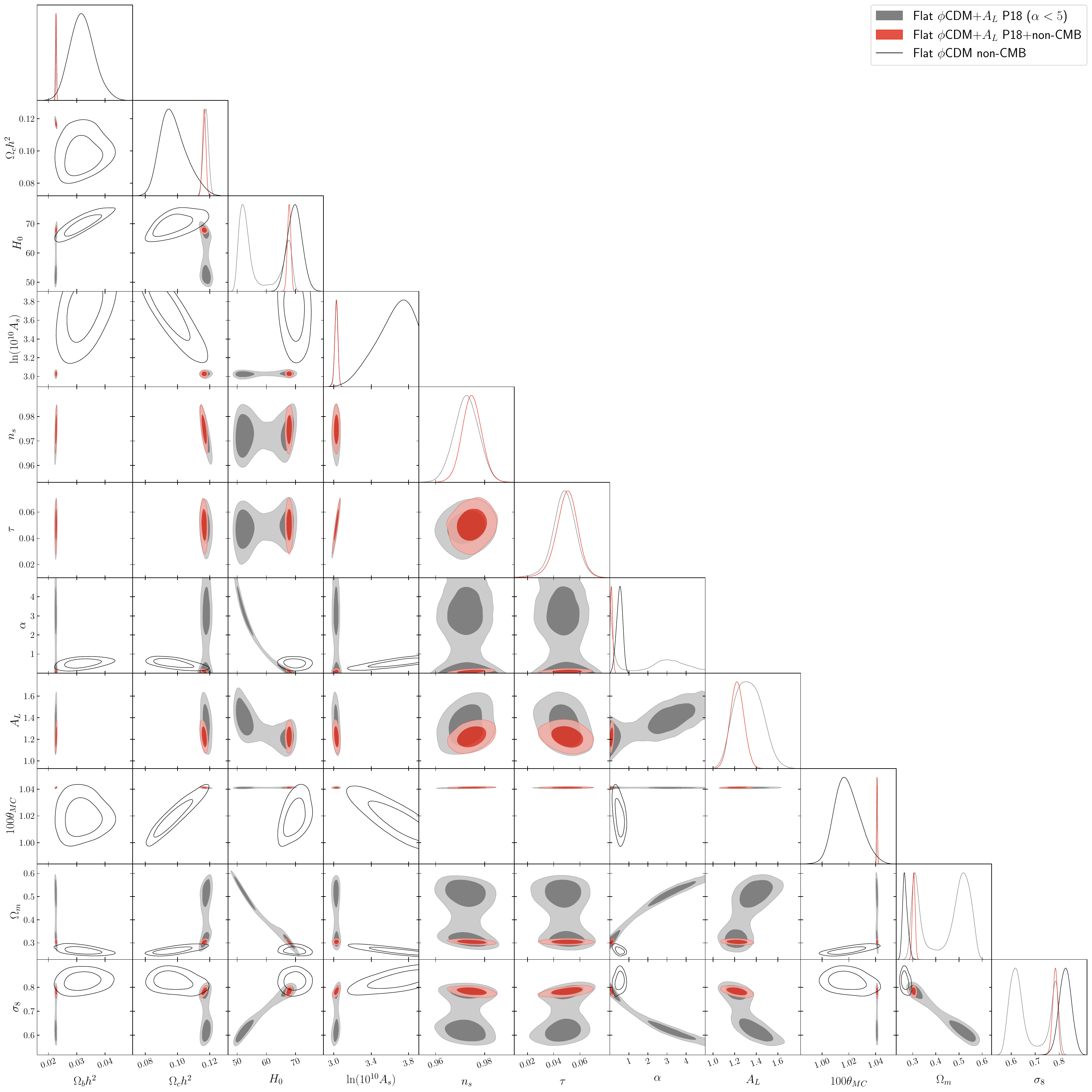}}
        \caption{One-dimensional likelihoods and 1$\sigma$ and $2\sigma$ likelihood confidence contours of flat $\phi$CDM$+A_L$ model parameters favored by non-CMB (solid curves), P18 (grey), and P18+non-CMB data sets (red contours). For P18 and P18+non-CMB cases, we include $\tau$ and $n_s$, which are fixed in the non-CMB data analysis. $H_0$ has units of km s$^{-1}$ Mpc$^{-1}$. For the P18 case the prior $\alpha \le 5$ was applied.
}
\label{fig:flat_phiCDM_Alens_P18_nonCMB23v2_a5_alensnstau}
\end{figure*}

%%%%%%%%%%%%%%%%%%%%%%%%%%%%%%%%%%%%%%%%%%%%%%%%%%%%%%%%%%%%%%%%%%%%%%%%%%%%%%%%%%%%%%%%%%%%%%%%
\begin{figure*}[htbp]
\centering
\mbox{\includegraphics[width=127mm]{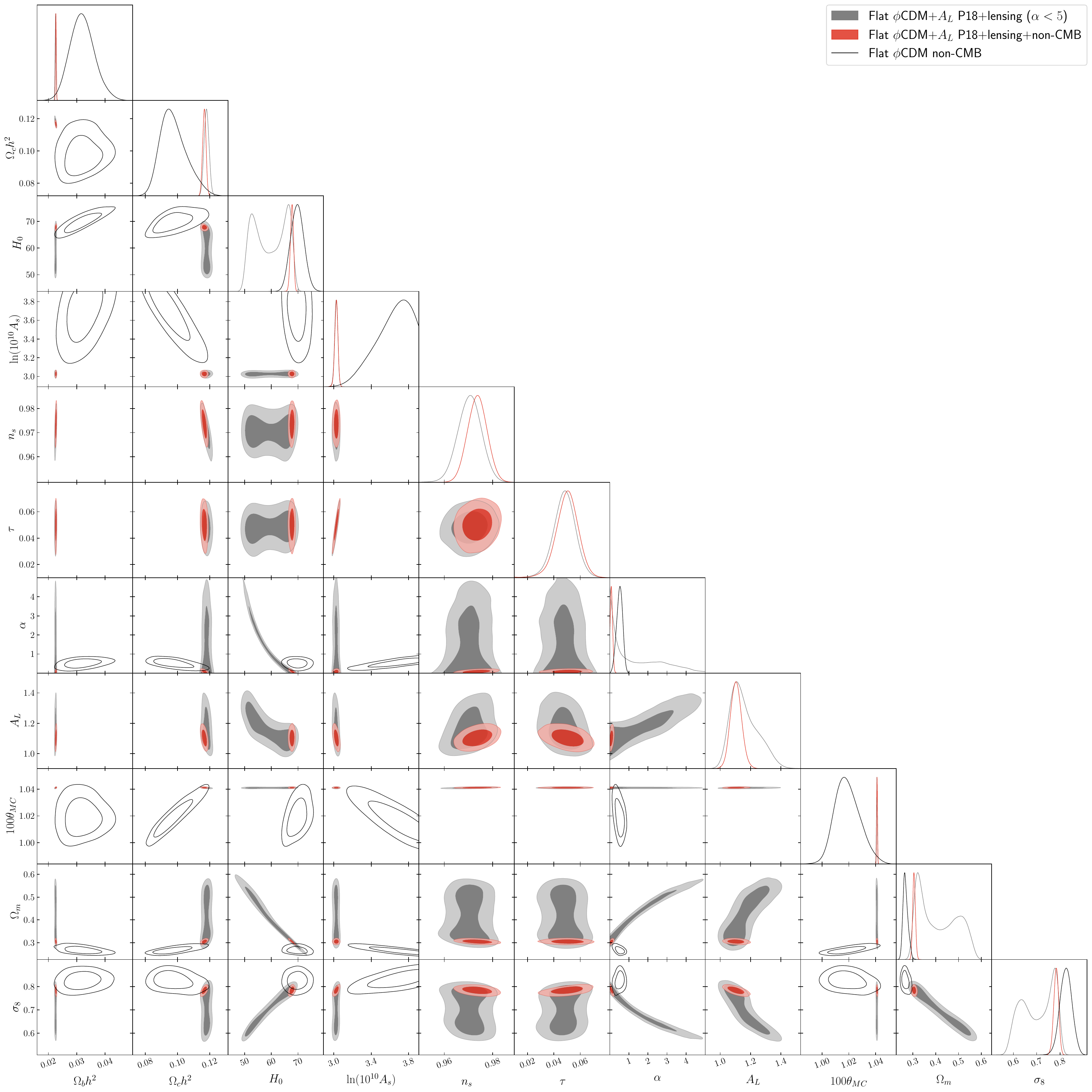}}
        \caption{One-dimensional likelihoods and 1$\sigma$ and $2\sigma$ likelihood confidence contours of flat $\phi$CDM$+A_L$ model parameters favored by non-CMB (solid curves), P18+lensing (grey), P18+lensing+non-CMB data sets (red contours). For P18+lensing and P18+lensing+non-CMB cases, we include $\tau$ and $n_s$, which are fixed in the non-CMB data analysis. $H_0$ has units of km s$^{-1}$ Mpc$^{-1}$. For the P18+lensing case the prior $\alpha \le 5$ was applied.
}
\label{fig:flat_phiCDM_Alens_P18_lensing_nonCMB23v2_a5_alensnstau}
\end{figure*}

%%%%%%%%%%%%%%%%%%%%%%%%%%%%%%%%%%%%%%%%%%%%%%%%%%%%%%%%%%%%%%%%%%%%%%%%%%%%%%%%%%%%%%%%%%%%%%%%
\begin{figure*}[htbp]
\centering
\mbox{\includegraphics[width=127mm]{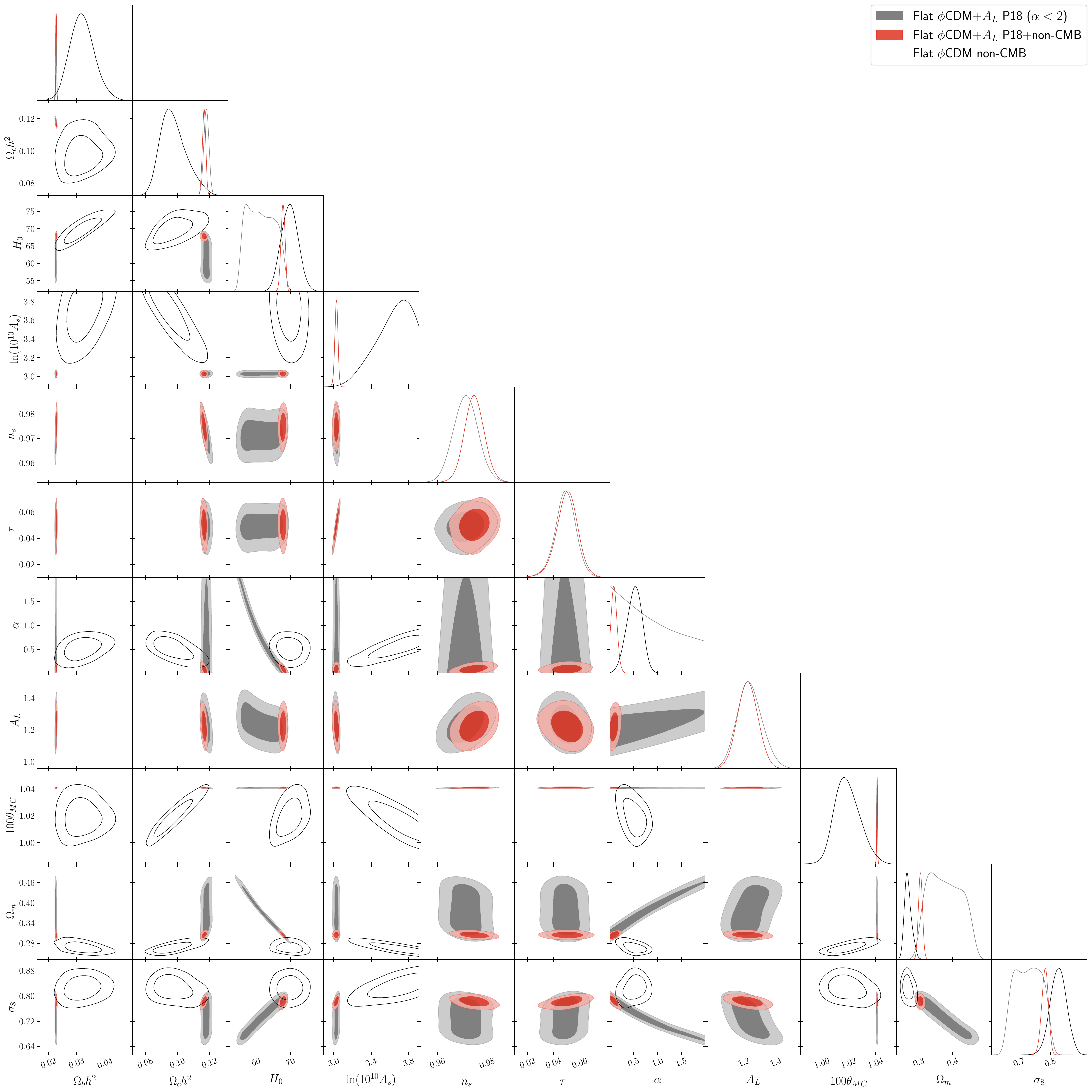}}
        \caption{One-dimensional likelihoods and 1$\sigma$ and $2\sigma$ likelihood confidence contours of flat $\phi$CDM$+A_L$ model parameters favored by non-CMB (solid curves), P18 (grey), and P18+non-CMB data sets (red contours). For P18 and P18+non-CMB cases, we include $\tau$ and $n_s$, which are fixed in the non-CMB data analysis. $H_0$ has units of km s$^{-1}$ Mpc$^{-1}$. For the P18 case the prior $\alpha \le 2$ was applied.
}
\label{fig:flat_phiCDM_Alens_P18_nonCMB23v2_a2_alensnstau}
\end{figure*}

%%%%%%%%%%%%%%%%%%%%%%%%%%%%%%%%%%%%%%%%%%%%%%%%%%%%%%%%%%%%%%%%%%%%%%%%%%%%%%%%%%%%%%%%%%%%%%%%
\begin{figure*}[htbp]
\centering
\mbox{\includegraphics[width=127mm]{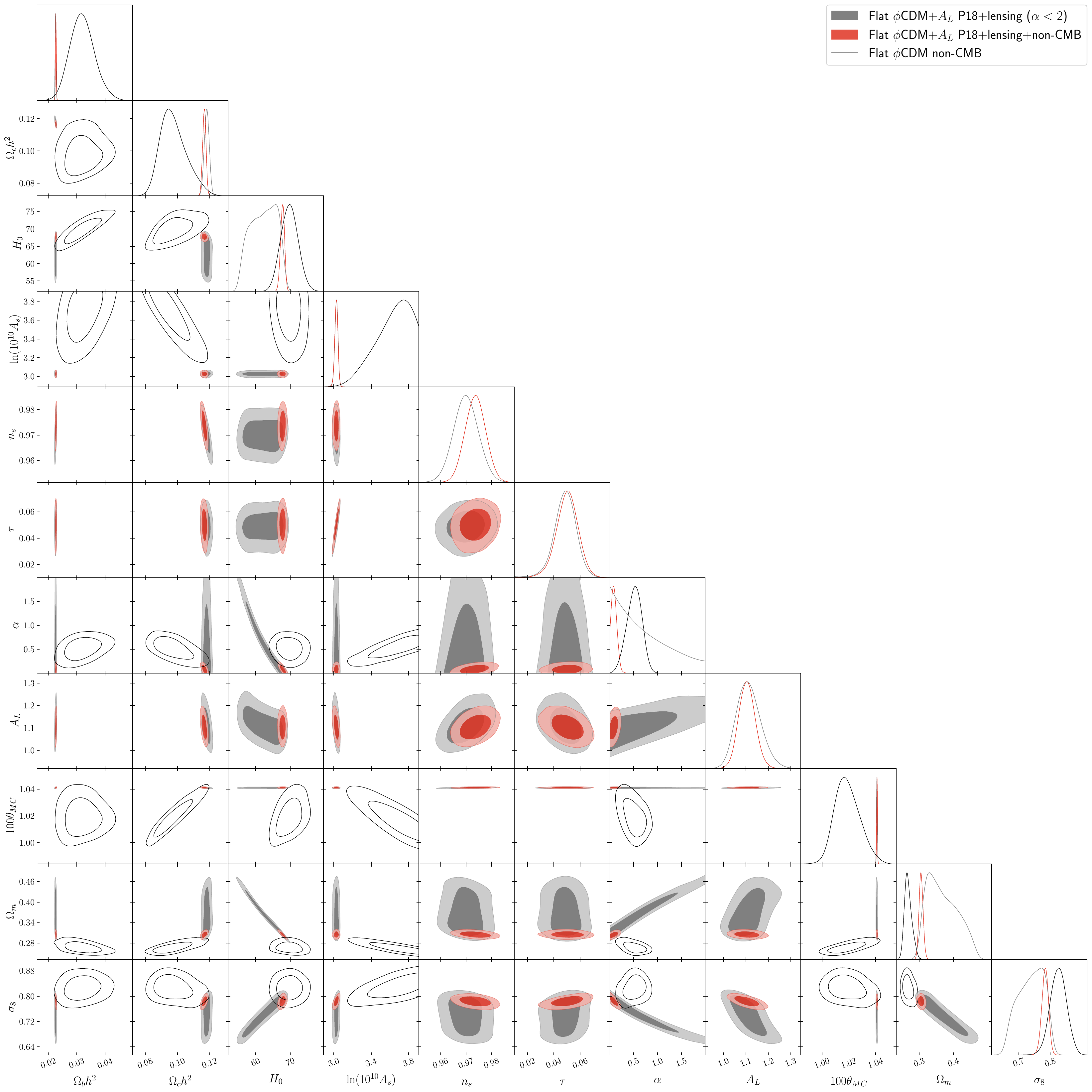}}
        \caption{One-dimensional likelihoods and 1$\sigma$ and $2\sigma$ likelihood confidence contours of flat $\phi$CDM$+A_L$ model parameters favored by non-CMB (solid curves), P18+lensing (grey), P18+lensing+non-CMB data sets (red contours). Fpr P18+lensing and P18+lensing+non-CMB cases, we include $\tau$ and $n_s$, which are fixed in the non-CMB data analysis. $H_0$ has units of km s$^{-1}$ Mpc$^{-1}$. For the P18+lensing case the prior $\alpha \le 2$ was applied.
}
\label{fig:flat_phiCDM_Alens_P18_lensing_nonCMB23v2_a2_alensnstau}
\end{figure*}

%%%%%%%%%%%%%%%%%%%%%%%%%%%%%%%%%%%%%%%%%%%%%%%%%%%%%%%%%%%%%%%%%%%%%%%%%%%%%%%%%%%%%%%%%%%%%%%
\begin{table}
\tbl{Consistency check parameter $\log_{10} \mathcal{I}$ and tension parameters $\sigma$ and $p$ for P18 vs.\ non-CMB data sets and P18+lensing vs.\ non-CMB data sets in the flat $\phi$CDM (+$A_L$) models. For the P18 and P18+lensing cases in the $\phi$CDM$+A_L$ model the prior $\alpha \le 2$ or $\alpha \le 5$ was applied.}
{\begin{tabular}{@{}lcc@{}} \toprule
                         & \multicolumn{2}{c}{Flat $\phi$CDM model}  \\[+1mm]
\cline{2-3}\\[-1mm]
   Data                          &  P18 vs non-CMB  & P18+lensing vs non-CMB  \\
\colrule
  $\log_{10} \mathcal{I}$        &   $-0.996$       & $-1.156$       \\
  $\sigma$                       &   $2.226$        & $2.540$        \\
  $p$ (\%)                       &   $2.601$        & $1.110$        \\
\colrule
                                 & \multicolumn{2}{c}{Flat $\phi$CDM$+A_L$ model ($\alpha \le 5$)}    \\[+1mm]
\cline{2-3}\\[-1mm]
  Data                           &   P18 vs non-CMB & P18+lensing vs non-CMB    \\
\colrule
  $\log_{10} \mathcal{I}$        &   $-1.023$       &  $-0.452$     \\
  $\sigma$                       &   $1.642$        &  $1.543$        \\
  $p$ (\%)                       &   $10.06$        &  $6.778$       \\
\colrule
                                 & \multicolumn{2}{c}{Flat $\phi$CDM$+A_L$ model ($\alpha \le 2$)}   \\[+1mm]
\cline{2-3}\\[-1mm]
  Data                           &   P18 vs non-CMB & P18+lensing vs non-CMB   \\
\colrule
  $\log_{10} \mathcal{I}$        &   $-0.610$       &  $-0.623$    \\
  $\sigma$                       &   $2.101$        &  $1.885$     \\
  $p$ (\%)                       &   $3.564$        &  $5.949$     \\
\botrule
\end{tabular}\label{tab:consistency_phiCDM}}
\end{table}
%%%%%%%%%%%%%%%%%%%%%%%%%%%%%%%%%%%%%%%%%%%%%%%%%%%%%%%%%%%%%%%%%%%%%%%%%%%%%%%%%%%%%%%%%%%%%%%

\section{Results and Discussion}
\label{sec:ResultsandDiscussion}

Cosmological parameter constraints are shown in Tables \ref{tab:results_flat_phiCDM}---\ref{tab:results_flat_phiCDM_a2} and in Figs.\ \ref{fig:flat_phiCDM_P18_nonCMB23v2_nstau}---\ref{fig:flat_phiCDM_Alens_P18_lensing_nonCMB23v2_a2_alensnstau}. Results obtained for the consistency between P18 and non-CMB and P18+lensing and non-CMB cosmological parameter constraints are displayed in Table \ref{tab:consistency_phiCDM}. The values of $\Delta \chi^2_{\text{min}}$, $\Delta$AIC and $\Delta$DIC, which are used to compare the performance of the flat $\Lambda$CDM model and the flat $\phi$CDM(+$A_L$) models, are listed in Tables \ref{tab:results_flat_phiCDM}---\ref{tab:results_flat_phiCDM_a2}.

Consistent with what we previously found when these data are analyzed using the XCDM, $w_0w_a$CDM, and $w(z)$CDM dynamical dark energy parameterizations (see Refs.\ \citen{deCruzPerez:2024abc, Park:2024vrw, Park:2024pew, Park:2025azv, deCruzPerez:2026mkg}), here when these data are analyzed using the physically consistent $\phi$CDM model the primary cosmological parameter related to the evolution of the dark energy, namely $\alpha$ in this case, is better constrained by the non-CMB data compilation considered than by either P18 or P18+lensing data.\footnote{In particular, when non-CMB data are analyzed in the context of the $\phi$CDM cosmological model, we find $\alpha = 0.52^{+0.17}_{-0.15}$, indicating a preference of $3.5\sigma$ for quintessence-like dark energy dynamics. This result is very similar to the phenomenon observed in the flat XCDM model constrained solely by non-CMB data, where the dark energy equation of state parameter $w=-0.853_{-0.033}^{+0.043}$ deviates by $4.5\sigma$ from the cosmological constant $w=-1$ and also favors quintessence-like dark energy dynamics.} This is because dark energy does not play a significant role at the higher redshift of CMB data. On the other hand, among the three derived parameters, non-CMB data are more effective than P18 or P18+lensing data at constraining only $\Omega_m$ in the $\phi$CDM ($+A_L$) models, as well as $\sigma_8$ in the $\phi$CDM$+A_L$ models, but, as expected, do not as effectively constrain $100\theta_{\rm MC}$.

In the $\phi$CDM+$A_L$ model, allowing $A_L$ to vary freely and adopting a wide $0 \le \alpha \le 10$ prior leads to bimodal likelihoods in the P18 and P18+lensing analyses, as already discussed in Sec.\ \ref{sec:Methods}. With $0 \le \alpha \le 5$, the bimodality persists, yielding slow but acceptable convergence ($R-1 < 0.0235$ for P18 and $R-1 < 0.0296$ for P18+lensing data).
Even if the MCMC does not satisfy the convergence criterion adopted here, the likelihood distributions and statistics for the parameters are sufficiently reliable if $R-1 < 0.1$ \cite{2018arXiv181209384V}. Figures \ref{fig:flat_phiCDM_Alens_P18_nonCMB23v2_a5_alensnstau} and \ref{fig:flat_phiCDM_Alens_P18_lensing_nonCMB23v2_a5_alensnstau} show the bimodality of the $0 \le \alpha \le 5$ prior results. The second peak near $\alpha = 3$, for the $0 \le \alpha \le 5$ prior case, is far from the part of parameter space favored by non-CMB data. We assume that the non-CMB measurements in our compilation are not grossly incorrect and so for the P18 and P18+lensing data analyses also consider a more restricted flat $\alpha$ prior non-zero only over $0 \le \alpha \le 2$, in which case the bimodality is mostly irrelevant, as can be seen in Figs.\ \ref{fig:flat_phiCDM_Alens_P18_nonCMB23v2_a2_alensnstau} and \ref{fig:flat_phiCDM_Alens_P18_lensing_nonCMB23v2_a2_alensnstau}, and $R-1 < 0.01$ convergence was achieved for the P18 and P18+lensing data sets. In the following we only focus on the $0 \le \alpha \le 2$ prior results for the $\phi$CDM+$A_L$ model P18 and P18+lensing data analyses.

Table \ref {tab:consistency_phiCDM} shows that non-CMB and P18 (P18+lensing) data constraints are incompatible at $2.2\sigma$ ($2.5\sigma$) in the flat $\phi$CDM model for the second, $p$ and $\sigma$, statistical estimator. The $\log_{10}\mathcal{I}$ estimator shows there is strong incompatibility between the two data sets in each pair, indicating that these results must be interpreted with caution. This should be compared to the $1.2\sigma$ ($1.2\sigma$) compatibility, $3.4\sigma$ ($3.6\sigma$) incompatibility, and $2.8\sigma$ ($2.7\sigma$) incompatibility between these two data sets in the flat $\Lambda$CDM model, the flat XCDM parameterization, and the flat $w_0w_a$CDM parameterization, respectively, see Tables X and XIV of Ref.\ \citen{deCruzPerez:2024abc} and Table 3 of Ref.\ \citen{Park:2024pew}, where according to $\log_{10}\mathcal{I}$ there is substantial compatibility (flat $\Lambda$CDM), decisive incompatibility (flat XCDM), substantial incompatibility (flat $w_0w_a$CDM), and here strong incompatibility (flat $\phi$CDM, Table \ref {tab:consistency_phiCDM}) between the two data sets in each pair. The results for the flat $\phi$CDM model lie between those of the XCDM and the $w_0w_a$CDM parameterizations, probably because $\phi$CDM cannot accommodate phantom-like dark energy dynamics while the other two can, and because $w_0w_a$CDM has one more free parameter than the other two. While it is possible to conclude that these incompatibilities between non-CMB and P18 data constraints and between non-CMB and P18+lensing data constraints rule out the flat $\phi$CDM model at $2.2\sigma$ and $2.5\sigma$ significance, these are below the $3\sigma$ threshold we adopt so we instead conclude that in the flat $\phi$CDM model non-CMB and P18 data and non-CMB and P18+lensing data are compatible at better than $3\sigma$ and can be jointly used to constrain cosmological parameters in this model. In the following we focus more on the P18+lensing+non-CMB data results, as that is the largest joint data set we study here. 

We have previously found that when the lensing consistency parameter $A_L$\cite{Calabreseetal2008} is also allowed to vary and be determined from these data, the incompatibilities between non-CMB and P18 (non-CMB and P18+lensing) data constraints are reduced \cite{deCruzPerez:2022hfr} to $0.16\sigma$ ($0.088\sigma$) compatibility, $2.1\sigma$ ($2.4\sigma$) incompatibility, and $1.9\sigma$ ($2.1\sigma$) incompatibility in the flat $\Lambda$CDM+$A_L$ model, the flat XCDM+$A_L$ parameterization, and the flat $w_0w_a$CDM+$A_L$ parameterization, respectively, see Tables X and XIV of Ref.\ \citen{deCruzPerez:2024abc} and Table 3 of Ref.\ \citen{Park:2024pew}. We find similar results for the flat $\phi$CDM+$A_L$ model here; from Table \ref{tab:consistency_phiCDM} we have $2.1\sigma$ ($1.9\sigma$) incompatibility between these two data sets for the $0 \le \alpha \le 2$ prior (and $1.6\sigma$ ($1.5\sigma$) incompatibility between these two data sets for the less-converged $0 \le \alpha \le 5$ prior results, possibly because in this case the likelihood bimodality discussed above results in larger error bars and so more compatible constraints), instead of the $2.2\sigma$ ($2.5\sigma$) incompatibility in the flat $\phi$CDM model with $A_L = 1$. From the $\log_{10} \mathcal{I}$ estimator we find substantial incompatibility in the $\phi$CDM+$A_L$ case for the $0 \le \alpha \le 2$ prior, Table \ref{tab:consistency_phiCDM}, but now reduced compared to the strong incompatibility in the $\phi$CDM case where $A_L = 1$, consistent with what we found for the flat $\Lambda$CDM+$A_L$, XCDM+$A_L$, and  $w_0w_a$CDM+$A_L$ models, see Tables X and XIV of Ref.\ \citen{deCruzPerez:2024abc} and Table 3 of Ref.\ \citen{Park:2024pew}.

Consistent with the numerical results shown in Table \ref {tab:consistency_phiCDM}, from Fig.\ \ref{fig:flat_phiCDM_P18_nonCMB23v2_nstau} (\ref{fig:flat_phiCDM_P18_lensing_nonCMB23v2_nstau}) we see that the $\phi$CDM model $2\sigma$ contours for non-CMB data and for P18 (P18+lensing) data have no overlap in the $\Omega_bh^2$---$\Omega_ch^2$, $\Omega_bh^2$---ln$(10^{10}A_s)$, $\Omega_ch^2$---ln$(10^{10}A_s)$, $\Omega_ch^2$---$H_0$, $H_0$---ln$(10^{10}A_s)$, and ln$(10^{10}A_s)$---$\alpha$ primary parameter subpanels. However, based on mean and $1\sigma$ confidence limits, unlike these $2\sigma$ contours, we can expect the $3\sigma$ contours to overlap.\footnote{Table \ref{tab:results_flat_phiCDM} compares results for the five-parameter flat $\phi$CDM model from non-CMB data with the seven-parameter version using P18 data. Significant differences in primary parameters are found for $\Omega_b h^2$ ($-2.1\sigma$), $\Omega_c h^2$ ($+3.6\sigma$), and $\ln(10^{10}A_s)$ ($-3.1\sigma$), while $\alpha$ differs by $-0.63\sigma$. For derived parameters, $100\theta_{\text{MC}}$, $\Omega_m$, and $\sigma_8$ have differences of $2.7\sigma$, $2.3\sigma$, and $-1.1\sigma$, respectively. Comparing P18+lensing to non-CMB results shows similar behaviors, with larger differences in $\Omega_b h^2$ ($-2.1\sigma$), $\Omega_c h^2$ ($+3.6\sigma$), and $\ln(10^{10}A_s)$ ($-3.1\sigma$). $\alpha$ differs by $-0.98\sigma$, while $100\theta_{\text{MC}}$, $\Omega_m$, and $\sigma_8$ differ by $2.7\sigma$, $2.4\sigma$, and $-1.1\sigma$.} These incompatibilities are somewhat reduced for the $\phi$CDM$+A_L$ model in Figs.\ \ref{fig:flat_phiCDM_Alens_P18_nonCMB23v2_a2_alensnstau} and \ref{fig:flat_phiCDM_Alens_P18_lensing_nonCMB23v2_a2_alensnstau} but the $2\sigma$ contours still have no overlap. The incompatibilities in these marginalized constraint contours seem to be more of a qualitative issue, whereas quantitative comparisons, such as the numerical $p$ and $\sigma$ values in Table \ref{tab:consistency_phiCDM}, are of greater importance.

From Table \ref{tab:results_flat_phiCDM} (\ref{tab:results_flat_phiCDM_a2}), and for P18+lensing+non-CMB data for the $\phi$CDM ($\phi$CDM$+A_L$) model, $\alpha = 0.055 \pm 0.041$ ($= 0.095 \pm 0.056$) differing from zero by $1.3\sigma$ ($1.7\sigma$), which appears to mildly favor quintessence-like dynamical dark energy over a cosmological constant. However, examining the likelihood contours reveals that $\alpha$ is most favored to be zero, with $\alpha < 0.196$ ($\alpha < 0.133$) being the 95\% confidence limits. Therefore, when P18 data are used, the $\phi$CDM model is consistent with the $\Lambda$CDM model, and a significant preference for dynamical dark energy only emerges when using non-CMB data alone. Also, as expected, from the same data compilation in the $\phi$CDM$+A_L$ model, from Table  \ref{tab:results_flat_phiCDM_a2}, $A_L = 1.105 \pm 0.037$ is $2.8\sigma$ larger than unity, a consequence of the observed excess smoothing of some of the P18 measured $C_\ell$'s. 

Comparing flat $\Lambda$CDM model cosmological parameter values determined from P18+ lensing+non-CMB data, given in the right column of the upper panel of Table IV of Ref.\ \citen{deCruzPerez:2024abc}, to those for the flat $\phi$CDM model from the same data compilation, given in the right column of Table  \ref{tab:results_flat_phiCDM} here, we find good agreement for the five common primary parameter values, with the differences being $-0.16\sigma$ for $\Omega_b h^2$, $0.34\sigma$ for $\Omega_c h^2$, $-0.18\sigma$ for $\tau$, $-0.19\sigma$ for $n_s$, and $-0.19\sigma$ for $\ln(10^{10}A_s)$, with also small differences for the four ``derived'' parameters, with $0.32\sigma$ for $100\theta_{\text{MC}}$, $0.77\sigma$ for $H_0$, $-0.50\sigma$ for $\Omega_m$, and $0.67\sigma$ for $\sigma_8$. It is reassuring that this data compilation provides cosmological parameter constraints that are almost independent of the assumed cosmological model.

Comparing flat $\Lambda$CDM$+A_L$ model cosmological parameter values determined from P18+ lensing+non-CMB data, given in the right column of Table VII of Ref.\ \citen{deCruzPerez:2024abc}, to those for the flat $\phi$CDM$+A_L$ model from the same data compilation, given in the right column of Table \ref{tab:results_flat_phiCDM_a2} here, we again find good agreement for the six common primary parameter values, with the differences being $-0.35\sigma$ for $\Omega_b h^2$, $0.74\sigma$ for $\Omega_c h^2$, $-0.20\sigma$ for $\tau$, $-0.47\sigma$ for $n_s$, $-0.12\sigma$ for $\ln(10^{10}A_s)$, and $-0.35\sigma$ for $A_L$, with also small, but a bit larger, differences for the four ``derived'' parameters, with $0.12\sigma$ for $100\theta_{\text{MC}}$, $0.99\sigma$ for $H_0$, $-0.59\sigma$ for $\Omega_m$, and $0.98\sigma$ for $\sigma_8$. 

Comparing flat $\phi$CDM model cosmological parameter values determined from P18+ lensing+non-CMB data, listed in the right column of Table \ref{tab:results_flat_phiCDM}, to those for the flat $\phi$CDM$+A_L$ model from the same data compilation, listed in the right column of Table \ref{tab:results_flat_phiCDM_a2}, for the seven primary parameters, gives differences of $-0.63\sigma$ for $\Omega_b h^2$, $1.1\sigma$ for $\Omega_c h^2$, $-0.22\sigma$ for $H_0$, $0.77\sigma$ for $\tau$, $-0.77\sigma$ for $n_s$, $0.98\sigma$ for $\ln(10^{10}A_s)$, and $-0.58\sigma$ for $\alpha$,\footnote{$\alpha$ is favored to be zero according to the 95\% upper limits, therefore, caution is required when interpreting the difference between the $\alpha$ values of the two models.} with differences for the three derived parameters of $-0.36\sigma$ for $100\theta_{\text{MC}}$, $0.55\sigma$ for $\Omega_m$, and $1.4\sigma$ for $\sigma_8$. The larger differences for $\Omega_c h^2$, $\ln(10^{10}A_s)$, and $\sigma_8$ are a consequence of the $2.8\sigma$ larger than unity value of $A_L = 1.105 \pm 0.037$ in the flat $\phi$CDM$+A_L$ model. 

From the P18+lensing+non-CMB data set in the flat $\phi$CDM$+A_L$ model we get $H_0=67.72^{+0.61}_{-0.54}$ km s$^{-1}$ Mpc$^{-1}$, which agrees with the median statistics result $H_0=68\pm 2.8$ km km s$^{-1}$ Mpc$^{-1}$ \cite{Chen:2011ab,Gottetal2001,Calabreseetal2012} as well as with some local measurements including the flat $\Lambda$CDM model value of Ref.\ \citen{Cao:2023eja} $H_0=69.25\pm 2.4$ km s$^{-1}$ Mpc$^{-1}$ from a joint analysis of $H(z)$, BAO, Pantheon+ SNIa, quasar angular size, reverberation-measured \mii\ and \civ\ quasar, and 118 Amati correlation gamma-ray burst data, and the local $H_0=69.03\pm 1.75$ km s$^{-1}$ Mpc$^{-1}$ from JWST TRGB+JAGB and SNIa data \cite{Freedman:2024eph}, but is in tension with the local $H_0=73.04\pm 1.04$ km s$^{-1}$ Mpc$^{-1}$ measured using Cepheids and SNIa data \cite{Riess:2021jrx}, also see Refs.\ \citen{Chen:2024gnu, Barua:2024gei}. Similarly, the flat $\phi$CDM$+A_L$ model with P18+lensing+non-CMB data yields $\Omega_m = 0.3052 \pm 0.0059$, which is in good agreement with the flat $\Lambda$CDM model value $\Omega_m = 0.313 \pm 0.012$ from Ref.\ \citen{Cao:2023eja} (based on the same data set described above for determining $H_0$).

From the $\Delta$DIC values in the last columns of Tables \ref{tab:results_flat_phiCDM} and \ref{tab:results_flat_phiCDM_a2} we see there is weak evidence for flat $\Lambda$CDM over flat $\phi$CDM and positive evidence for flat $\phi$CDM$+A_L$ over flat $\Lambda$CDM.

\section{Conclusion}
\label{sec:Conclusion}

We have tested the spatially flat dynamical dark energy $\phi$CDM($+A_L$) cosmological model, without and with a variable lensing consistency parameter $A_L$, with different combinations of CMB and non-CMB data. We find that the scalar field parameter $\alpha$, which governs dark energy dynamics, is more tightly constrained by non-CMB data than by CMB data alone. Since the non-CMB data results and the P18+lensing data results are compatible at better than $3\sigma$ we can use these data together in a joint analysis. From an analysis of this largest of the data sets we use, P18+lensing+non-CMB data, we obtain $\alpha = 0.055 \pm 0.041$ ($\alpha <0.133$, 95\% upper limit) in the $\phi$CDM model and $\alpha = 0.095 \pm 0.056$ ($\alpha < 0.196$, 95\% upper limit) in the $\phi$CDM+$A_L$ model, both of which are consistent with a cosmological constant ($\alpha=0$), but allow mild quintessence-like dark energy dynamics at $1.3\sigma$ and $1.7\sigma$.
We again note that unlike what we find in the XCDM, $w_0w_a$CDM, and $w(z)$CDM parameterizations analyzed with the same data compilation used here, \cite{deCruzPerez:2024abc, Park:2024pew, Park:2025azv} in the $\phi$CDM model when $A_L$ is allowed to vary and be measured from these data the support for dark energy dynamics increases, compared to the $A_L = 1$ case.

The estimated Hubble constant is $H_0=67.55_{-0.46}^{+0.53}$ km s$^{-1}$ Mpc$^{-1}$ from P18+lensing+non-CMB data in the $\phi$CDM model, consistent with median statistics and some local determinations, but in tension with other local determinations. The constraints for the non-relativistic matter density and the clustering amplitude ($\Omega_m = 0.3096 \pm 0.0055$, $\sigma_8 = 0.8013_{-0.0067}^{+0.0077}$) in the flat $\phi$CDM model are statistically consistent with those in the $\Lambda$CDM model. Allowing the CMB lensing amplitude consistency parameter $A_L$ to vary reduces tensions between CMB data and non-CMB data constraints, although we find $A_L = 1.105 \pm 0.037$, $2.8\sigma$ higher than unity, consistent with the excess smoothing seen in Planck data. 

AIC and DIC model comparisons show that, for these data, the $\phi$CDM model provides a fit comparable to the $\Lambda$CDM model, with the $\phi$CDM+$A_L$ model extension slightly preferred in some cases. Overall, our results indicate that while the $\Lambda$CDM model remains an excellent fit, current data leave open the possibility of mildly evolving quintessence-like dynamical dark energy, and so do not require dynamical dark energy with a phantom-divide crossing. Future, more precise observations will be essential for distinguishing between the cosmological constant and the dynamical evolution predicted by other physically consistent models such as $\phi$CDM.

\section*{Acknowledgments}

C.-G.P.\ was supported by the National Research Foundation of Korea (NRF) grant funded by the Korea government (MSIT) No.\ RS-2026-25473390.

\bibliographystyle{ws-ijmpd}
\bibliography{PhiCDM}

\end{document}